# Similarity between quantum mechanics and thermodynamics: Entropy, temperature, and Carnot cycle

Sumiyoshi Abe[1,2,3] and Shinji Okuyama[1]

[1] *Department of Physical Engineering, Mie University, Mie 514-8507, Japan*
[2] *Institut Supérieur des Matériaux et Mécaniques Avancés, 44 F. A. Bartholdi, 72000 Le Mans, France*
[3] *Inspire Institute Inc., Alexandria, Virginia 22303, USA*

**Abstract**

Similarity between quantum mechanics and thermodynamics is discussed. It is found that if the Clausius equality is imposed on the Shannon entropy and the analogue of the quantity of heat, then the value of the Shannon entropy comes to formally coincide with that of the von Neumann entropy of the canonical density matrix, and pure-state quantum mechanics apparently transmutes into quantum thermodynamics. The corresponding quantum Carnot cycle of a simple two-state model of a particle confined in a one-dimensional infinite potential well is studied, and its efficiency is shown to be identical to the classical one.

In their work [1], Bender, Brody, and Meister have developed an interesting discussion about a quantum-mechanical analogue of the Carnot engine. They have treated a two-state model of a single particle confined in a one-dimensional potential well with width $L$ and have considered reversible cycle. Using the "internal" energy, $E(L) = \langle \psi | H | \psi \rangle$, they define the pressure (i.e., the force, because of the single dimensionality) as $f = -dE(L)/dL$, where $H$ is the system Hamiltonian and $|\psi\rangle$ is a quantum state. Then, the analogue of "isothermal" process is defined to be $fL = $ const. corresponding to the classical equation of state of the ideal gas at fixed temperature, whereas the analogue of "adiabatic" process is characterized by $fL^3 = $ const., which is the Poisson-like law. The efficiency of the cycle they obtained reads

$$\eta = 1 - \frac{E_C}{E_H}, \qquad (1)$$

where $E_H$ ($E_C$) is the value of $E$ along the "isothermal" process at high (low) "temperature".

This discovery is remarkable, since it shows how the work can be extracted from a simple quantum-mechanical system through particular deformations of the quantum state as well as the potential width.

However, our point here is to grasp the above discussion as a hidden similarity between quantum mechanics and thermodynamics at a certain level. Then, it is equally interesting to examine if it is possible to fully establish such similarity. To do so, of central importance is to identify the analogues of the entropy and temperature. According to Kelvin, the efficiency of the Carnot cycle can be used to define the absolute thermodynamic temperature as $E_C / E_H = T_C / T_H$ [2]. Thus, an analogue of the "law of equipartition of energy", which is violated in quantum theory, is assumed in the Carnot cycle in Ref. [1]. Therefore, a question that naturally arises then is if this "temperature" is consistent with the concept of "entropy".

In this paper, we revisit this intriguing problem of similarity between quantum mechanics and thermodynamics, not by requiring the analogies of the classical equation of state, equipartition of energy, the Poisson law, etc., but by imposing only the Clausius equality as a fundamental thermodynamic relation. Then, we show that the value of the Shannon entropy becomes formally equal to that of the von Neumann entropy of the canonical density matrix, and pure-state quantum mechanics apparently transmutes into equilibrium quantum thermodynamics [3]. However, decay of a pure state into a mixed state itself never appears explicitly. We shall also discuss, just for comparison with the result in Ref. [1], the corresponding Carnot cycle in a simple two-state model of a particle confined in a one-dimensional infinite potential well. Its efficiency is found to be in complete agreement with the classical one.

Consider a quantum-mechanical system in a state $|\psi\rangle$. The mean value of energy to be identified with the analogue of the "internal" energy is

$$E = \langle \psi | H | \psi \rangle, \tag{2}$$

where $H$ is the system Hamiltonian. Under changes of both the Hamiltonian and the

state, $E$ varies as

$$\delta E = \left(\delta\langle\psi|\right)H|\psi\rangle + \langle\psi|\delta H|\psi\rangle + \langle\psi|H\left(\delta|\psi\rangle\right). \tag{3}$$

This expression has a formal analogy with the first law of thermodynamics

$$\delta'Q = \delta E + \delta'W, \tag{4}$$

if the following correspondence relations are noticed for the changes of the analogues of the quantity of heat and work:

$$\delta'Q \leftrightarrow \left(\delta\langle\psi|\right)H|\psi\rangle + \langle\psi|H\left(\delta|\psi\rangle\right), \tag{5}$$

$$\delta'W \leftrightarrow -\langle\psi|\delta H|\psi\rangle, \tag{6}$$

respectively.

To further elaborate the analogy between quantum mechanics and thermodynamics, it is necessary to identify the concepts of entropy and temperature. Regarding the entropy, one might naively imagine that the entropy relevant to quantum thermodynamics may be the von Neumann entropy

$$S^{vN} = -k\,\mathrm{Tr}(\rho\ln\rho). \tag{7}$$

As known well, however, this is not the case, since it identically vanishes in the case when the density matrix, $\rho$, is of a pure state. Instead, the authors of Ref. [1] suggest the use of the Shannon entropy. Using the complete orthonormal system, $\{|u_n\rangle\}_n$, which are the eigenstates of the Hamiltonian, $H$, satisfying $H|u_n\rangle = E_n|u_n\rangle$ with $\{E_n\}_n$ being the energy eigenvalues, we can expand a general state, $|\psi\rangle$, as follows:

$$|\psi\rangle = \sum_n a_n |u_n\rangle, \tag{8}$$

$$\sum_n |a_n|^2 = 1. \tag{9}$$

Then, the Shannon entropy is given in terms of the expansion coefficients as

$$S^S = -k \sum_n |a_n|^2 \ln |a_n|^2. \tag{10}$$

$S^{vN}$ is invariant under unitary transformations of $\rho$, whereas $S^S$ is not invariant under unitary transformations of $|\psi\rangle$, in general. That is, $S^S$ depends on the choice of a basis. The reason why the energy eigenbasis is employed here is as follows. In the subsequent thermodynamic-like discussion, a crucial role is played by $S^S$, which should not exhibit explicit dynamical time evolution. The expansion coefficients, $a_n$'s, in terms of the energy eigenbasis dynamically change in time as follows: $a_n \to a_n \exp(-iE_n t/\hbar)$. Therefore, $|a_n|^2$ remains unchanged, leading to time independence of $S^S$, as desired. It is also noted that the positive constant, $k$, appearing here is still not necessarily the Boltzmann constant in this stage.

However, it turns out to be impossible to establish the Clausius-like equality for reversible processes

$$\delta S^S = \frac{\delta' Q}{T}, \qquad (11)$$

within the framework of Ref. [1]. Thus, similarity between quantum mechanics and thermodynamics is not yet complete in the quantum Carnot cycle.

Now, our idea is to impose Eq. (11) *to define the analogue of temperature*. To do so, first we rewrite Eq. (2) as follows:

$$E = \sum_n E_n |a_n|^2. \qquad (12)$$

Since we are interested in the thermodynamic-like situation, the time scale associated with the change of the state is much larger than that of a dynamical one, $\sim \hbar / E$. In this case, the adiabatic theorem [4] applies. The change of the state in Eq. (5), i.e., the analogue of the quantity of heat, is given by

$$\delta' Q = \sum_n E_n \delta |a_n|^2. \qquad (13)$$

Regarding Eqs. (12) and (13), see the later discussion above Eq. (17). On the other hand, the change of the Shannon entropy reads

$$\delta S^S = -k \sum_n \left( \ln |a_n|^2 \right) \delta |a_n|^2. \tag{14}$$

A point of crucial importance is that if the Clausius-like equality in Eq. (11) is imposed on Eqs. (13) and (14), then the expansion coefficient becomes the following "canonical form":

$$|a_n|^2 = \frac{1}{Z} e^{-\frac{E_n}{kT}}, \tag{15}$$

where $Z = \sum_n \exp(-E_n / kT)$. Accordingly, the Shannon entropy is calculated to be $S^S = -k \sum_n \left[ \exp(-E_n/kT)/Z \right] \ln \left[ \exp(-E_n/kT)/Z \right]$, which precisely coincides with the value of the von Neumann entropy in Eq. (7) of the "canonical density matrix",

$$\rho = \frac{1}{Z} e^{-\frac{H}{kT}} \tag{16}$$

with $Z = \text{Tr} \exp(-H/kT)$. This implies that imposition of the Clausius equality transmutes pure-state quantum mechanics into genuine quantum thermodynamics, if $k$ and $T$ are regarded as the Boltzmann constant and real temperature, respectively.

Based on this fact, let us discuss as in Ref. [1] the Carnot cycle of a simple two-state model of a single particle confined in a one-dimensional infinite potential well with width $L$, in order to compare pure-state quantum mechanics with quantum thermodynamics. An arbitrary state, $|\psi\rangle$, is given by superposition of the ground and first excited states, $|u_1\rangle$ and $|u_2\rangle$: $|\psi\rangle = \sum_{n=1,2} a_n |u_n\rangle$. $L$ varies much slower compared with the dynamical time scale, $\sim \hbar/E$, as mentioned earlier. The system Hamiltonian, $H$, its eigenstate, $|u_n\rangle$, and the corresponding energy eigenvalue, $E_n$, all depend on changing $L$. Write them as $H(L)$, $|u_n(L)\rangle$, and $E_n(L)$, respectively. They satisfy the equation, $H(L)|u_n(L)\rangle = E_n(L)|u_n(L)\rangle$, in the adiabatic approximation. This justifies Eqs. (12) and (13). The expectation value of the Hamiltonian in this state, $|\psi\rangle$, is given by

$$E = \frac{\pi^2 \hbar^2}{2mL^2}\xi + \frac{4\pi^2 \hbar^2}{2mL^2}(1-\xi), \qquad (17)$$

where $m$ is the mass of the particle and $\xi \equiv |a_1|^2 \in [0,1]$ (and thus, $|a_2|^2 = 1-\xi$). The change of this quantity through the variation of $L$ contains two contributions, as in Eq. (3). As repeatedly emphasized, the time scale of this variation is assumed to be much larger than the dynamical one, and so the change of the state is represented by the change of $\xi$. Therefore, as in Eq. (13), the change of the analogue of the quantity of heat is

$$\delta' Q = -\frac{3\pi^2 \hbar^2}{2mL^2}\delta\xi. \qquad (18)$$

On the other hand, the change of the work in Eq. (6) in the present case is calculated to be

$$\delta'W = \frac{\pi^2 \hbar^2}{mL^3}(4-3\xi)\,\delta L. \qquad (19)$$

Also, the change of the Shannon entropy reads

$$\delta S^S = k\ln\!\left(\frac{1}{\xi}-1\right)\delta\xi. \qquad (20)$$

From Eqs. (18) and (20), it is seen that identification of the "internal" energy with "temperature" is not consistent.

Here, we introduce $T$ through Eq. (11). Accordingly, $\xi$ is solved as a function of $L$ and $T$ as follows:

$$\xi(L,T) = \frac{1}{1+\exp\!\left(-\dfrac{3\pi^2\hbar^2}{2mL^2 kT}\right)}. \qquad (21)$$

Note that $1/2 < \xi(L,T) < 1$, which guarantees that the probability of finding the system in the ground state is always larger than that in the excited state, as desired. Now, let us consider the Carnot cycle (Fig. 1): $A \to B \to C \to D \to A$, where the processes $A \to B$ and $C \to D$ are the isothermal processes with high- and low-temperature, $T_H$ and $T_C$, respectively, whereas $B \to C$ and $D \to A$ are the adiabatic processes with fixed $\xi$. The values of the volume (i.e., width) at A, B, C, and D are $L_1$, $L_2$, $L_3$, and $L_4$. The heat absorbed by the system is calculated from Eq. (18) as

$$Q_H = \int_{(A)}^{(B)} d'Q = -\frac{3\pi^2\hbar^2}{2m}\int_{L_1}^{L_2} dL\, \frac{1}{L^2}\frac{d\xi(L,T_H)}{dL}$$

$$= \frac{3\pi^2\hbar^2}{2m}\left[\frac{1-\xi(L_2,T_H)}{L_2^2} - \frac{1-\xi(L_1,T_H)}{L_1^2}\right] + kT_H \ln\frac{\xi(L_1,T_H)}{\xi(L_2,T_H)}. \tag{22}$$

On the other hand, using Eq. (19), we can calculate the work during each process as follows:

$$W_{A\to B} = \int_{(A)}^{(B)} d'W = \int_{L_1}^{L_2} dL\, \frac{\pi^2\hbar^2}{mL^3}\left[4 - 3\xi(L,T_H)\right]$$

$$= \frac{\pi^2\hbar^2}{2m}\left(\frac{1}{L_1^2} - \frac{1}{L_2^2}\right) + kT_H \ln\frac{\xi(L_1,T_H)}{\xi(L_2,T_H)}, \tag{23}$$

$$W_{B\to C} = \int_{(B)}^{(C)} d'W = \int_{L_2}^{L_3} dL\, \frac{\pi^2\hbar^2}{mL^3}\left[4 - 3\xi(L_2,T_H)\right]$$

$$= \frac{\pi^2\hbar^2}{2m}\left[4 - 3\xi(L_2,T_H)\right]\left(\frac{1}{L_2^2} - \frac{1}{L_3^2}\right), \tag{24}$$

$$W_{C\to D} = \int_{(C)}^{(D)} d'W = \int_{L_3}^{L_4} dL\, \frac{\pi^2\hbar^2}{mL^3}\left[4 - 3\xi(L,T_C)\right]$$

$$= \frac{\pi^2\hbar^2}{2m}\left(\frac{1}{L_3^2} - \frac{1}{L_4^2}\right) + kT_C \ln\frac{\xi(L_3,T_C)}{\xi(L_4,T_C)}, \tag{25}$$

$$W_{D \to A} = \int_{(D)}^{(A)} d'W = \int_{L_4}^{L_1} dL \frac{\pi^2 \hbar^2}{mL^3}\left[4 - 3\xi(L_4, T_C)\right]$$

$$= \frac{\pi^2 \hbar^2}{2m}\left[4 - 3\xi(L_4, T_C)\right]\left(\frac{1}{L_4^2} - \frac{1}{L_1^2}\right). \tag{26}$$

Using the equalities associated with the adiabatic processes, $B \to C$ and $D \to A$,

$$\xi(L_2, T_H) = \xi(L_3, T_C), \tag{27}$$

$$\xi(L_4, T_C) = \xi(L_1, T_H), \tag{28}$$

that is,

$$L_2^2 T_H = L_3^2 T_C, \tag{29}$$

$$L_4^2 T_C = L_1^2 T_H, \tag{30}$$

we find that the total work done by the cycle is given by

$$W = W_{A \to B} + W_{B \to C} + W_{C \to D} + W_{D \to A}$$

$$= \left(1 - \frac{T_C}{T_H}\right) Q_H \tag{31}$$

with $Q_H$ given in Eq. (22). Therefore, we obtain the efficiency of the cycle as follows:

$$\eta = 1 - \frac{T_C}{T_H}, \tag{30}$$

which precisely coincides with that of the classical Carnot cycle.

In conclusion, we have studied a structural similarity between quantum mechanics and thermodynamics. We have focused our attention on the Shannon entropy in the energy eigenbasis and the analogue of the quantity of heat. We have found that if the Clausius equality is imposed on them, then the value of the Shannon entropy becomes that of the von Neumann entropy of the canonical density matrix, and pure-state quantum mechanics apparently transmutes into quantum thermodynamics. We have, however, recognized that decay of a pure state into a mixed state itself does not explicitly appear in the discussion. To examine the result in comparison with the work in Ref. [1], we have also studied the Carnot cycle of a two-state model of a particle confined in a one-dimensional infinite potential well and have seen that its efficiency is identical to the classical one.

S. A. was supported in part by a Grant-in-Aid for Scientific Research from the Japan Society for the Promotion of Science.

---

Figure Caption

FIG. 1 The Carnot cycle depicted in the plane of the width (*L*) and force (*f*). The processes, $A \to B$ ($C \to D$) and $B \to C$ ($D \to A$), are isothermal expansion at high temperature (compression at low temperature) and adiabatic expansion (compression), respectively. The cycle can in fact be realized with the functional function form of $\xi(L,T)$ in Eq. (21).

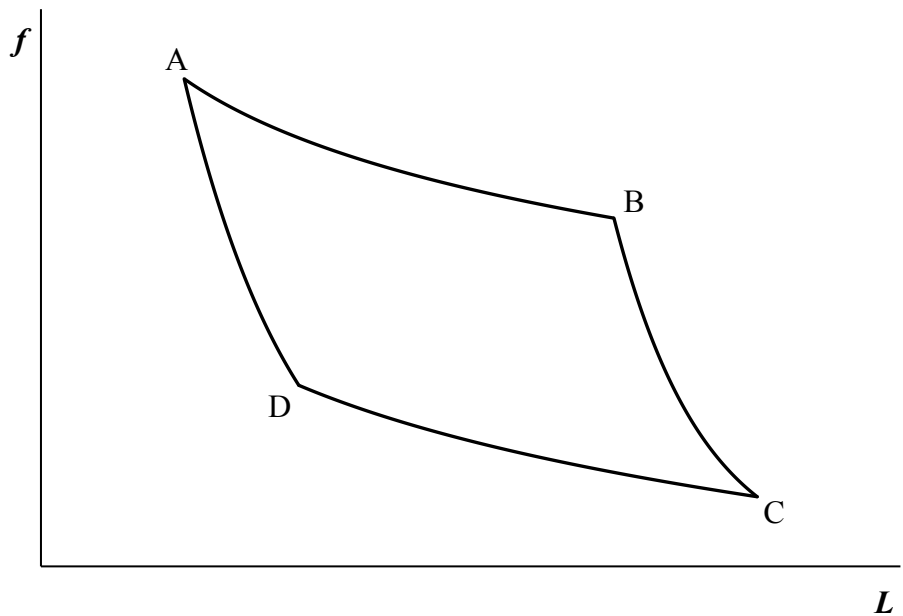

Figure 1